\documentclass[aps,pra,reprint,superscriptaddress]{revtex4-1}
\usepackage{bbold}
\usepackage{amsmath,amsfonts,amsmath,mathtools,bbm,bm}
\usepackage{graphicx}
\usepackage{braket}
\usepackage[colorlinks,linkcolor=blue,urlcolor=blue,citecolor=blue]{hyperref}
\usepackage[all]{hypcap}
\usepackage[normalem]{ulem}

\usepackage{blindtext}
\usepackage{enumitem}
\usepackage{xcolor}

\setcitestyle{numbers,square}

\newcommand{\Tr}{\mathrm{Tr}}

\def\({\left (}
\def\){\right )}

\begin{document}

\title{Extensive Multipartite Entanglement from su(2) Quantum Many-Body Scars}

\author{Jean-Yves Desaules} 
 \affiliation{School of Physics and Astronomy, University of Leeds, Leeds LS2 9JT, United Kingdom}

\author{Francesca Pietracaprina}
\affiliation{Department of Physics, Trinity College Dublin, Dublin 2, D02PN40, Ireland}

\author{Zlatko Papi\'c}
\email{z.papic@leeds.ac.uk}
\affiliation{School of Physics and Astronomy, University of Leeds, Leeds LS2 9JT, United Kingdom}

\author{John Goold}
\email{gooldj@tcd.ie}
\affiliation{Department of Physics, Trinity College Dublin, Dublin 2, D02PN40, Ireland}

\author{Silvia Pappalardi}
\email{silvia.pappalardi@phys.ens.fr}
\affiliation{Laboratoire de Physique de l'\'Ecole Normale Sup\'erieure, ENS, Universit\'e PSL, CNRS,  Sorbonne Universit\'e, Universit\'e de Paris, F-75005 Paris, France}

\date{\today}

\begin{abstract}
Recent experimental observation  of weak ergodicity breaking  in Rydberg atom quantum simulators has sparked interest in quantum many-body  scars --  eigenstates which evade thermalisation at finite energy densities due to novel mechanisms  that do not rely on integrability or protection by a global symmetry.  A salient feature of some quantum many-body scars is their sub-volume bipartite entanglement entropy. In this work we demonstrate that such exact many-body scars also possess extensive \emph{multipartite} entanglement structure if they stem from an su(2) spectrum generating algebra. We show this analytically, through scaling of the quantum Fisher information, which is found to be super-extensive for exact scarred eigenstates in contrast to generic thermal states. Furthermore, we numerically study signatures of multipartite entanglement in the PXP model of Rydberg atoms, showing that extensive quantum Fisher information density can be generated dynamically by performing a global quench experiment. Our results identify a rich multipartite correlation structure of scarred states with significant potential as a resource in quantum enhanced metrology.
\end{abstract}

\maketitle

\label{sec:intro}
 
{\bf Introduction}---  Beyond fundamental importance in quantum information theory~\cite{family}, entanglement now plays  a central role in many-body physics~\cite{Amico2008, Laflorencie2016,  DeChiara18}. For example, the finite-size scaling of bipartite entanglement  allows one to deduce important information on  critical scalings in many-body systems~\cite{Eisert2010}, and through the identification of area and volume law behaviour it can tell us about the feasibility of classical simulation. Entanglement is also central to the foundations of statistical mechanics~\cite{popescu2006}. Due to advances in experimental ultra-cold atomic physics, significant effort has been made to understand how quantum systems thermalise in the long-time limit~\cite{Rigol:2008,Alessio:2016,eisert2015quantum}. Thermalising systems which obey the eigenstate thermalisation hypothesis (ETH)~\cite{Srednicki:1994,Srednicki:1996,Srednicki:1999,Alessio:2016}  have eigenstates that obey a volume law entanglement entropy. This agrees with common intuition that the highly-excited energy eigenstates of many body systems are close to random vectors in the Hilbert space and as such should be highly entangled~\cite{Page93}. Volume-law bipartite entanglement entropy, therefore, is ubiquitous in nature. 
Unfortunately, this structure of bipartite entanglement is not known to lead to any practical advantage for quantum-enhanced technologies. 

There are many facets to entanglement theory and, in particular, many-body systems offer the perfect playground to explore multipartite entanglement. 
In this work we focus on the multipartite structure in the eigenstates of complex many-body systems as described by the quantum Fisher Information (QFI) ~\cite{helstrom1969quantum, Tth2014, pezze2014quantum}. The latter quantifies the usefulness of the the quantum state as a resource for quantum enhanced metrology and can be directly related to multipartite entanglement~\cite{Hyllus2012,Tth2012,Pezze2009}.  One particularly appealing feature of QFI is its relation to thermal susceptibilities~\cite{Hauke2016, pappalardi2017multipartite, Brenes2020}, hence its experimental accessibility in condensed matter physics~\cite{strobel2014fisher, Hauke2016}. 
In fact, when computed for a thermal canonical Gibbs state 
the QFI can be written directly in terms of a Kubo response function \cite{Hauke2016}. This has led to experiments with neutron scattering \cite{laurell2021quantifying, scheie2021witnessing} and experimental proposals in atomic platforms \cite{deAlmeida2021entanglement}.
One may then ask the question: are there eigenstates of local many-body systems with an entanglement structure that could have operational significance for quantum information processing? 

In this work, we demonstrate that systems with weak ergodicity breaking, possessing eigenstates known as quantum many-body scars (QMBS)~\cite{turner2018weak, wenwei18TDVPscar}, naturally realise  a non-trivial form of  extensive multipartite entanglement. 
QMBS are ETH-violating eigenstates which span a subspace that is effectively decoupled from the thermalising bulk of the spectrum of a non-integrable many-body system~\cite{serbyn2021, PapicReview, MoudgalyaReview}.  
Such subspaces have been shown to arise due to several complementary mechanisms, including 
Hilbert space fragmentation~\cite{Khemani2019_2,Sala2019,MoudgalyaKrylov,bosonScars, Zhao2020} and projector embedding~\cite{ShiraishiMori,NeupertScars,Wildeboer2020,Surace2020}.
In this article, we focus on a large class of QMBS  arising due to su(2) spectrum generating algebras~\cite{BernevigEnt, Iadecola2019_2, mark_unified_2020, MoudgalyaHubbard, Iadecola2019_3, OnsagerScars, Chattopadhyay,Lee2020,Pakrouski2020, Ren2020, Dea2020, comparin2021universal}. The latter mechanism gives rise to an extensive number of QMBS eigenstates with equal energy spacing, leading to robust quantum  revivals, usually from a simple product state. 
The su(2) QMBS are naturally realized in the so-called PXP spin model~\cite{TurnerPRB, Iadecola2019, Lin2019, Shiraishi_2019, Lin2020, MondragonShem2020, turner2020correspondence}, in which signatures of QMBS have been observed experimentally~\cite{bernien2017, Su2022}.
While previous studies of QMBS have focused extensively on their bipartite entanglement, demonstrating an ETH violation via the sub-volume law entanglement entropy, the study of \emph{multipartite} entanglement of QMBS has so far been lacking.  We show that exact su(2) QMBS have QFI that scales super-extensively with system size, in contrast to generic thermal states of locally interacting Hamiltonians. Furthermore, remnants of this non-trivial scaling can be detected using dynamical quench experiments in systems with approximate QMBS, as we demonstrate numerically using the PXP model of Rydberg atoms~\cite{turner2018weak}. As the QFI is related to the well-known response functions in condensed matter physics~\cite{Hauke2016,pappalardi2017multipartite,Brenes2020}, this provides an opportunity for detection of multipartite entanglement in experiment and applications in quantum-enhanced sensing~\cite{dooley}.  

{\bf{Quantum Fisher information and multipartite entanglement}}---
The QFI, $\mathcal F_Q$, is a central concept in quantum metrology that sets ultimate bounds on the precision on the estimation of a parameter~\cite{SM}.
 The general goal is to estimate an unknown parameter $\lambda$ using a quantum state $\hat \rho$. By performing a quantum measurement protocol, one finds the precision is constrained by the quantum Cram\'er-Rao bound $(\Delta \lambda)^2 \geq 1/M \mathcal F_Q(\hat \rho_\lambda)$, where $M$ is the number of independent measurements made in the protocol, $\hat \rho_\lambda$ is the state parametrized by $\lambda$ and $\Delta \lambda$ is the variance \cite{Braunstein1994,Pezze2018}.
The $\mathcal F_Q$ admits an exact expression when the state $\hat{\rho}_{\lambda}$ is generated by some Hermitian operator $\hat{O}$ such that $\hat \rho_{\lambda}= e^{i\lambda \hat O}\hat \rho  e^{-i\lambda \hat O}$. For a general mixed state, described by the density matrix $\hat \rho=\sum_n p_n |n\rangle\langle n|$,  it reads \cite{Braunstein1994}
\begin{eqnarray}
\label{full_QFI}
{\mathcal F_Q}(\hat O, \hat \rho){=}2\sum_{n,m} \frac{(p_n-p_{m})^2}{p_n+p_{m}} |\langle n| \hat O|m\rangle|^2 {\leq 4 \, \langle \Delta \hat O^2 \rangle } ,
\end{eqnarray}
with $\langle  \Delta \hat O^2 \rangle =\text{Tr}(\hat \rho\, \hat O^2 ) -  \text{Tr}(\hat \rho\,\hat O )^2$. The equality holds for pure states $\hat \rho = \ket{\psi}\bra{\psi}$.

The QFI has key mathematical properties \cite{Braunstein1994,PETZ2011,Tth2014, Pezze2018}, such as convexity, additivity, monotonicity, and it
can be used to probe the multipartite entanglement structure of a quantum state~\cite{Hyllus2012,Tth2012,Pezze2009}. If, in a system with $N$ particles and for a certain \emph{collective} operator $\hat O = \frac 12 \sum_{i=1}^N \hat o_i$ (extensive sums of operators $\hat o_i$ with local support), the QFI density satisfies the inequality
\begin{equation}
\label{eq:qfiD}
 f_Q \equiv \frac{\mathcal{F}_Q(\hat O, \hat \rho)}{N} > m  \ ,
\end{equation}
then, at least $(m+1)$ parties in the system are entangled (with $1\leq m \leq N-1$ a divisor of $N$). Namely, $m$ represents the size of the biggest entangled block of the quantum state.
In particular, if $N-1 \leq f_Q(\hat O) \leq N$, then the state is called \emph{genuinely $N$-partite entangled}. 
\vspace{.1cm}

{\bf QFI of thermal eigenstates}---
In general, different operators $\hat{O}$ lead to different bounds on QFI and there is no systematic method (without some knowledge on the physical system \cite{Hauke2016,Pezz2017Gabri}) to choose the optimal one. 
In this work, we restrict ourselves to one-dimensional systems and collective operators $\hat O = \frac 12 \sum_{i=1}^N \hat o_i$, which are typically explored in cold-atoms experiments and in interferometric schemes \cite{Pezze2018}. 
For the eigenstates $\ket{E_n}$, the QFI with respect to such collective operators $\mathcal F_Q(\hat O, \ket{E_n}) = 4{\bra{E_n} \Delta \hat O^2\ket{E_n}}$ can be expressed in terms of the connected correlation functions $G_{i,j}(E_n) \equiv \langle E_n|{ \hat o_i \hat o_{j}}|E_n\rangle -   \langle E_n|{ \hat o_i}|E_n\rangle  \langle E_n|{ \hat o_{j}}|E_n\rangle$.
If we further assume \emph{translational invariance}, then $G_{i,j}=G_{|i-j|}$ and the QFI density \eqref{eq:qfiD} reads
\begin{equation}
\label{eq:QFICONNE}
f_Q(\hat O, \ket{E_n}) =  G_0(E_n) + 2\sum_{r=1}^{N-1} G_r(E_n) \ .
\end{equation}
 Note that $G_0(E_n)=\mathcal O(1)$ is always an intensive quantity \footnote{ For instance, $G(E_n)=1$ for spin operators and eigenstates in the middle of the spectrum.}, hence the scaling of $f_Q$ depends on the behavior of $G_r(E_n)$ as a function of the distance $r$.

We now study the scaling of the QFI density \eqref{eq:QFICONNE} for generic chaotic eigenstates of a locally-interacting many-body Hamiltonian far from criticality, which are well-known to obey ETH \cite{Srednicki:1999}. In this case, the connected correlation functions scale as
\begin{equation}
\label{eq:conneCh}
G_r(E_n) \sim c_r e^{-r/\xi} \ , \quad r \gg \xi \ ,
\end{equation}
where $|c_r|=\mathcal O(1)$ is an intensive constant that depends on the operators and $\xi$ is the correlation length at energy $E_n$. 
This is a consequence of the clustering property of connected correlation functions of local observables, which has been demonstrated for canonical thermal states \cite{araki1969gibbs}. Appealing to ETH \cite{Garrison2018}, the same clustering property holds for eigenstates of local hamiltonians 
up to sub-extensive corrections \cite{kuwahara2020eigenstate}. The decay of correlations in Eq.~\eqref{eq:conneCh} holds despite the volume-law entanglement entropy of the eigenstates~\cite{Deutsch2013Microscopic, Vidmar2017Entanglement, Murthy2019Structure}, see the discussion in~\cite{SM} and Refs.~\cite{Iadecola2019, qi2019determining} for numerical examples. 

By plugging Eq.~\eqref{eq:conneCh} into Eq.~\eqref{eq:QFICONNE} and summing over $r$, we obtain for $N\gg 1$
\begin{align}
\begin{split}
\label{eq:qfiChaos}
	f_Q(\hat O, \ket{E_n}) 
	& \lesssim G_0(E_n) + \frac{2c}{e^{1/\xi}-1} + \mathcal O(e^{-N})	\ ,
\end{split}
\end{align}
where we have used $|c_r|\leq c=\mathcal O(1)$. This equation shows that generically the QFI density of chaotic eigenstates, away from criticality, is an intensive quantity that can be evaluated explicitly from the knowledge of the thermal correlation length. Furthermore,  whenever the correlation length  $\xi$ is large (but finite), one has
\begin{equation}
f_Q(\hat O, \ket{E_n})  \simeq 2  \xi  \quad \text{for}\quad \xi \gg 1 \ .
\end{equation}
Thus, the QFI is also large and finite. By comparing this expression with the relation to multipartite entanglement \eqref{eq:qfiD}, we find that the size of the biggest entangled block scales as twice the correlation length. This finding is fully consistent with known results for critical pure or thermal states, where the QFI for the order parameter diverges universally~\cite{Hauke2016, Brenes2020, zanardi2008quantum, gabbrielli2018multipartite,
frerot2018quantum, frerot2019reconstructing}. 

As a side note, the above result for pure chaotic eigenstates satisfying ETH can be compared with that for QFI of thermal states \cite{Hauke2016} (or the asymptotic state of a quenched dynamics). In the latter case, the QFI bounds from above the corresponding canonical expression the Gibbs state ${\cal F}(\hat{O}, \ket{E})\ge {\cal F}(\hat{O}, \hat \rho_{\textrm{Gibbs}})$ \cite{Brenes2020}.

{\bf{QFI of exact scars}}---
We next contrast the scaling of the QFI for thermal eigenstates~\eqref{eq:qfiChaos} to the one for a class of \emph{exact} QMBS. More precisely, we focus our work on scarred eigenstates that can be described within the general framework of Mark, Lin and Motrunich~\cite{mark_unified_2020} (see also Ref.~\onlinecite{MoudgalyaHubbard}). Whenever there exists a linear subspace $W\subset \mathcal H$ of the Hilbert space and an operator $\hat Q^\dagger$ such that $\hat Q^\dagger W \subset W$ and 
\begin{equation}
	\label{eq:ds}
	\left  ( [\hat H, \hat Q^\dagger] -\omega \hat Q^\dagger \right ) W = 0 \ ,
\end{equation}
then the Hamiltonian admits the following exact eigenstates $\ket{\mathcal S_n}$ and corresponding eigenvalues $E_n$, 
\begin{align}
\label{eq:es}
\ket{\mathcal S_n} = (\hat Q^{\dagger})^n \ket{\mathcal S_0} \ ,
\quad  E_n = E_0 + n\omega\ ,
\end{align}
where $\ket{\mathcal S_0}$ is an eigenstate of the Hamiltonian $\hat H$ with eigenvalue $E_0$. In other words, $\hat Q^\dagger$ is a dynamical symmetry of the Hamiltonian restricted to the subspace $W$. The specific form of the operator $\hat Q^\dagger$ is model dependent. Typically, it is a collective operator with momentum $\pi$, e.g.,  in one dimension $	\hat Q^\dagger = \sum_{i=1}^N(-1)^i\, \hat o_i$ with $\hat o_i$ an operator with local support~\cite{mark_unified_2020}. Note that Eq.~\eqref{eq:es} implies equal energy spacing amongst the scarred eigenstates, and so any state that would have overlap only on these states would show perfect wavefunction revivals.
Let us define
\begin{align}
\label{eq:defineJs}
\hat 	 J^+ \equiv \frac {\hat Q^\dagger}2 \ ,\quad
\hat 	 J^- \equiv \frac {\hat Q}2 \ ,\quad
\hat 	 J^z \equiv \frac {\hat H}\omega \ ,
\end{align}
which forms the Cartan-Weyl basis of an su(2) algebra.
We will use the following notation $\hat A=_w \hat B$ meaning that the equality holds only on the subset $W$ \eqref{eq:es}. For instance, Eq.~\eqref{eq:ds} reads
$	[\hat J^z, \hat J^{\pm}] =_w \pm\hat  J^{\pm}$.

Depending on how the algebra is completed, one may obtain different results on the scaling of correlations. If, for instance, one has $[\hat J^+, \hat J^-]_w=1$ -- the standard algebra of the harmonic oscillator -- then $\hat J^{\pm}$ act like creation and annihilation operators, while $\hat J^+ \hat J^-$ acts as a number operator. It follows
\begin{equation}
	\frac{\bra{\mathcal S_n} \hat J^+ \hat J^- \ket{\mathcal S_n}}{N^2}  = 
	\frac 1N \frac nN \ ,
\end{equation}
and there cannot be any long-range order. Suppose, instead, that the operators $\hat J$ obey 
\begin{equation}
    \label{eq:SU2}
	[\hat J^+, \hat J^{-}] =_w 2\hat  J^z \ .
\end{equation}
For such an algebraic structure one can show~\cite{SM}
\begin{equation}
	\label{eq:lro}
	\frac{\bra{\mathcal S_n} \hat J^+\hat  J^- \ket{\mathcal S_n}}{N^2} 
	= \frac{2\epsilon_0}{\omega} \frac{n}{N} - 
	\left(\frac{n}{N}\right)^2 + \frac{n}{N^2}\ ,
\end{equation}
where $\epsilon_0$ is the the ground state energy density, i.e. ${E_0 = -N\epsilon_0}$. As $n=0$ to $N$, the first two terms are $\mathcal{O}(1)$ while the last one is only $\mathcal{O}(1/N)$.

Hence, exact scars with finite energy density ($n/N=\mathcal O(1)$) possess \emph{long-range order}~\cite{yang1962concept}.  
As such, for the local operators $\hat o_i$ appearing in $\hat Q^\dagger$, the connected correlation functions are finite in the thermodynamic limit \eqref{eq:LRCorre}, i.e.
\begin{equation}
	\label{eq:LRCorre}
	G_r(E_n) \sim \text{const} \ , \quad r\to \infty \ , \quad  N\to \infty \ .
\end{equation}
This property was used in Ref.~\cite{Iadecola2019} to interpret scarred eigenstates as finite-energy-density condensates of weakly interacting $\pi$-magnons that possess long-range order in both space and time.
A key results of our findings is that, through Eq.~\eqref{eq:QFICONNE}, the presence of long-range order implies genuine multipartite entanglement of this class of QMBS. In fact, the QFI density with respect to the operators $\hat J^x=(\hat J^++ \hat J^-)/2$ reads
\begin{equation}\label{eq:fq_su2}
	f_Q(\hat J^x, \ket{\mathcal S_n}) = 2 \left( \frac{2 \epsilon_0}{\omega} - \frac {n}{N}\right )n + \frac{2 \epsilon_0}{\omega} \ ,  
\end{equation}
where we used  $\bra{\mathcal S_n} \hat J^x\ket{\mathcal S_n}=\bra{\mathcal S_n}(\hat J^{\pm})^2\ket{\mathcal S_n}=0$ to get rid of all terms except the ones in Eq.~\eqref{eq:lro}. Therefore exact scars with finite energy density $n\sim N$ possess super-extensive QFI $F_Q \sim N^2$ and they are genuinely multipartite entangled. In general, it is highly non-trivial to engineer super- extensive scaling of quantum Fisher information for many-body states ~\cite{PETZ2011}. The identification of such states as a subspace in the spectrum of physical, locally-interacting systems is our central result.

\begin{figure}[t]
	\centering
	\includegraphics[width=\linewidth]{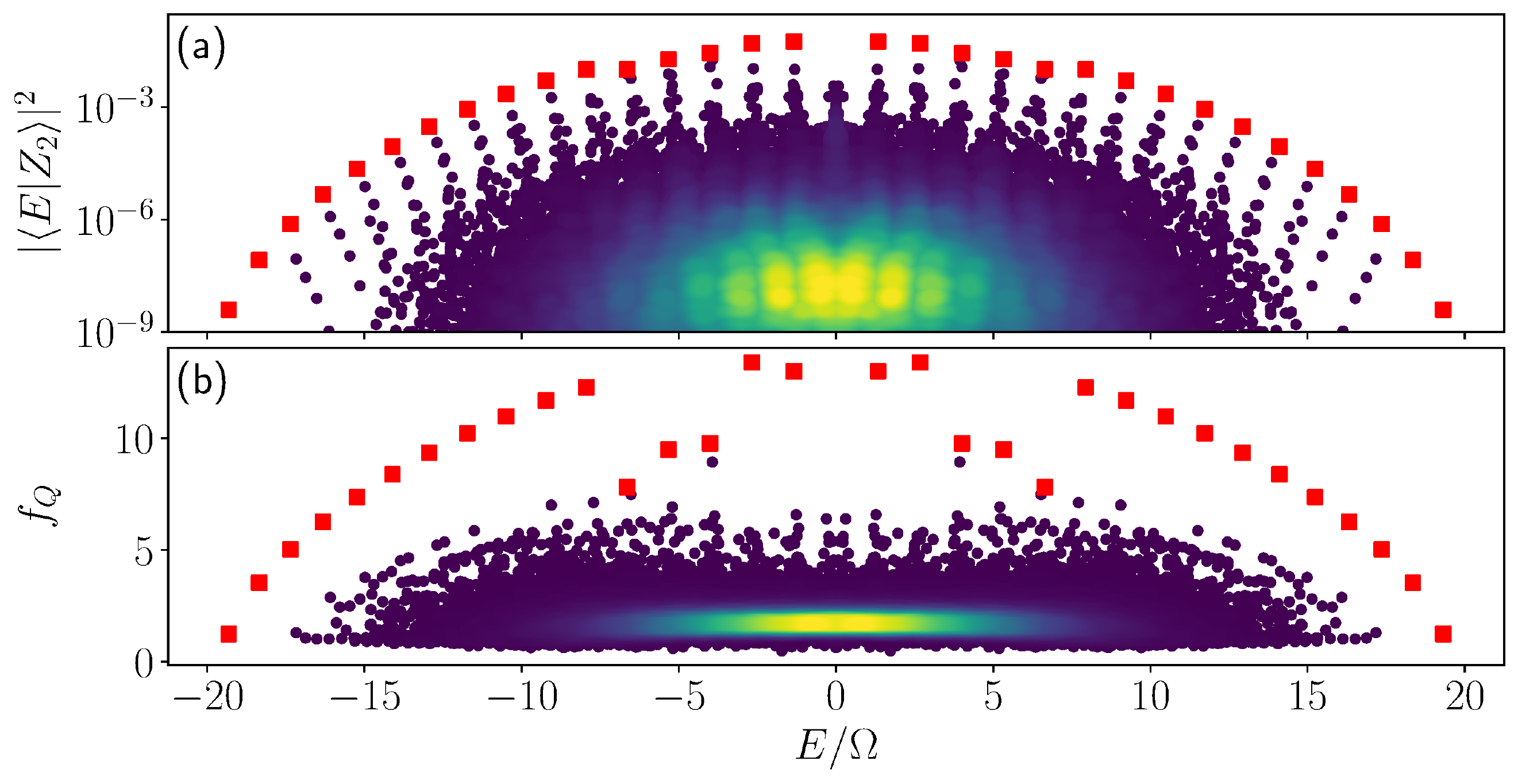}
	\caption{\small (a) Overlap between exact PXP eigenstates and the N\' eel state. Red squares indicate the QMBS eigenstates. (b) The QFI density of the PXP eigenstates. The red squares denote the same QMBS eigenstates as in (a). In both plots, the dips in the middle of the spectrum are due to hybridisation of QMBS eigenstates with thermal states. The colour code indicates the density of points and all data is for the PXP model in Eq.~\eqref{eq:H_PXP} with $N{=}32$ spins.}
	\label{fig:PXP_QFI_eigs}
\end{figure}
{\bf{Experimental implications for Rydberg atoms.}}--- Signatures of QMBS have been observed in experiments on Rydberg-blockaded atomic chains~\cite{bernien2017} and Bose-Hubbard quantum simulators~\cite{Su2022}. Denoting by $\ket{{\circ}}$ and $\ket{{\bullet}}$ the ground and excited states of each atom, respectively, the effective ``PXP" Hamiltonian describing such systems is given by~\cite{Fendley2004,Lesanovsky2012} 
\begin{equation}\label{eq:H_PXP}
    \hat{H}=\Omega\sum_j \hat P_{j-1} \hat X_j \hat P_{j+1},
\end{equation}
where $\hat X{=}\ket{{\circ}}\bra{{\bullet}}+\ket{{\bullet}}\bra{{\circ}}$ is the Pauli operator, $\hat P{=}\ket{{\circ}}\bra{{\circ}}$ is the projector on the ground state of an atom, and $\Omega$ is the Rabi frequency. We also assume periodic boundary conditions. The Hamiltonian in Eq.~\eqref{eq:H_PXP} is compatible with the Rydberg blockade constraint as it allows an atom to change state only if both of its neighbours are in the ground state, thus neighbouring excitations such as $\ket{\cdots {\bullet}{\bullet}\cdots}$ are excluded.

While the model in Eq.~\eqref{eq:H_PXP} is non-integrable and thermalising~\cite{turner2018weak}, when quenched from the N\' eel product state of atoms, $\ket{\mathbb{Z}_2}=\ket{{\bullet}{\circ}{\bullet}{\circ}\cdots}$, long-lived oscillations are seen in the dynamics  of entanglement entropy and local observables~\cite{bernien2017,turner2018weak,TurnerPRB}. 
This is despite $\ket{\mathbb{Z}_2}$ state being effectively at ``infinite-temperature''.  The origin of these oscillations is a set of $N{+}1$ QMBS eigenstates~\cite{turner2018weak} that form an approximate su(2) algebra ~\cite{Choi2018,Bull2020}. However, this algebra is only approximate, hence Eq.~\eqref{eq:ds} is not exactly obeyed; moreover, the algebra involves non-local generators, hence it is not easy to directly measure. To circumvent this problem, we employ the alternating magnetic field operator,
\begin{equation}\label{eq:MS}
\hat M_S=\frac{1}{2}\sum_j(-1)^j \hat Z_j,
\end{equation}
where $\hat Z{=}\ket{{\bullet}}\bra{{\bullet}}-\ket{{\circ}}\bra{{\circ}}$.
This operator is natural because it is experimentally accessible  and it is proportional to the total spin $\hat J^x$ operator in the approximate su(2) algebra of Ref.~\onlinecite{Iadecola2019}, while QMBSs are eigenstates of the corresponding $\hat J^z$ operator, defined via Eq.\eqref{eq:SU2}.

\begin{figure}[t]
	\centering
	\includegraphics[width=\linewidth]{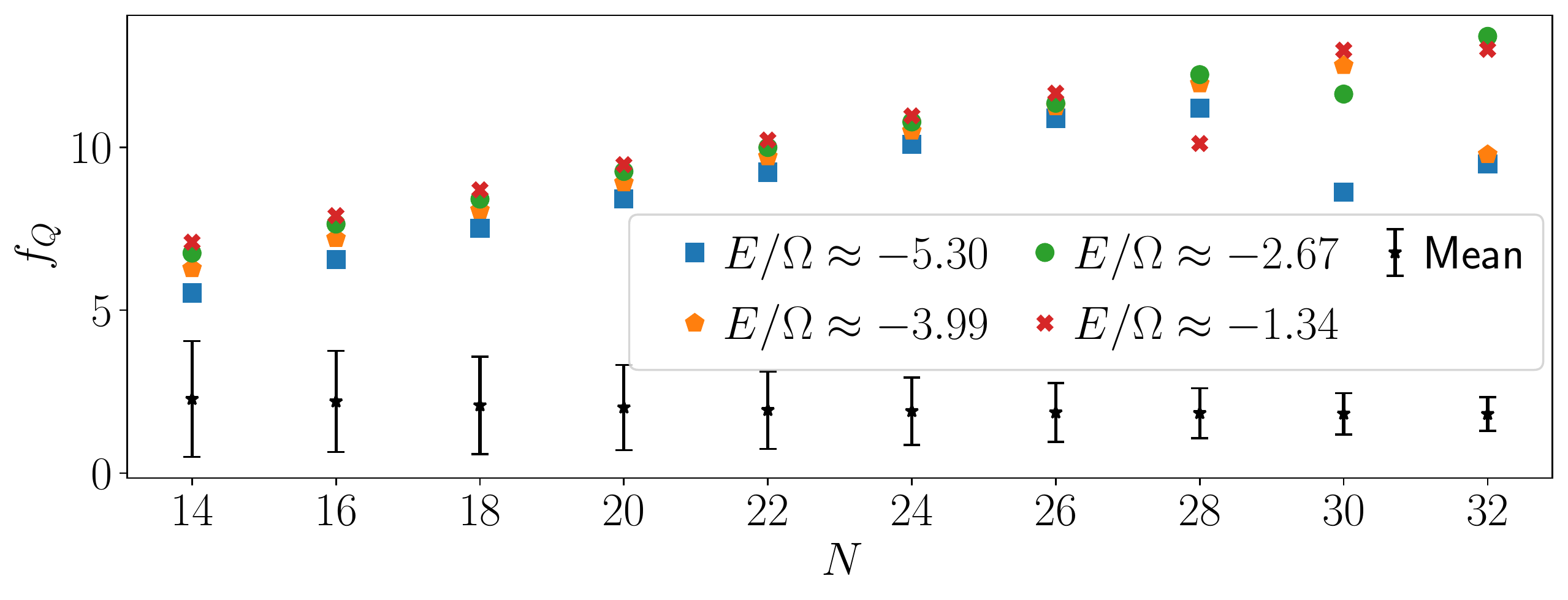}
	\caption{\small Finite size scaling of QFI density for several QMBS eigenstates of the PXP model with energies $E$ near the middle of the spectrum, contrasted against the mean value over all eigenstates. The scaling is extensive until $N{=}28$, where hybridisation between the scarred eigenstates and thermal eigenstates with a similar energy starts to lower $f_Q$.}
	\label{fig:PXP_QFI_scaling}
\end{figure}

Fig.~\ref{fig:PXP_QFI_eigs} shows that, as for exact scars, the QMBS eigenstates in the PXP model have largest QFI among all eigenstates. 
Further differences in the connected correlation functions between QMBS and other thermalising eigenstates are also observed~\cite{SM}. However, as the scarred PXP subspace is weakly connected to the rest of the Hilbert space, in  larger systems the QMBS eigenstates begin to hybridise with thermal eigenstates~\cite{TurnerPRB}, which is manifested as a reduction in QFI and the overlap with the N\'eel state. Signatures of this in the middle of the spectrum can be observed in Fig.~\ref{fig:PXP_QFI_eigs}. Hybridisation also prevents the QFI of the individual QMBS states to scale super-extensively beyond a certain size, see  Fig.~\ref{fig:PXP_QFI_scaling}. The same figure also shows that, for thermal eigenstates, $f_Q$ does not depend on $N$, as predicted in Eq.~\eqref{eq:qfiChaos}. 

\begin{figure}[t]
	\centering
	\includegraphics[width=\linewidth]{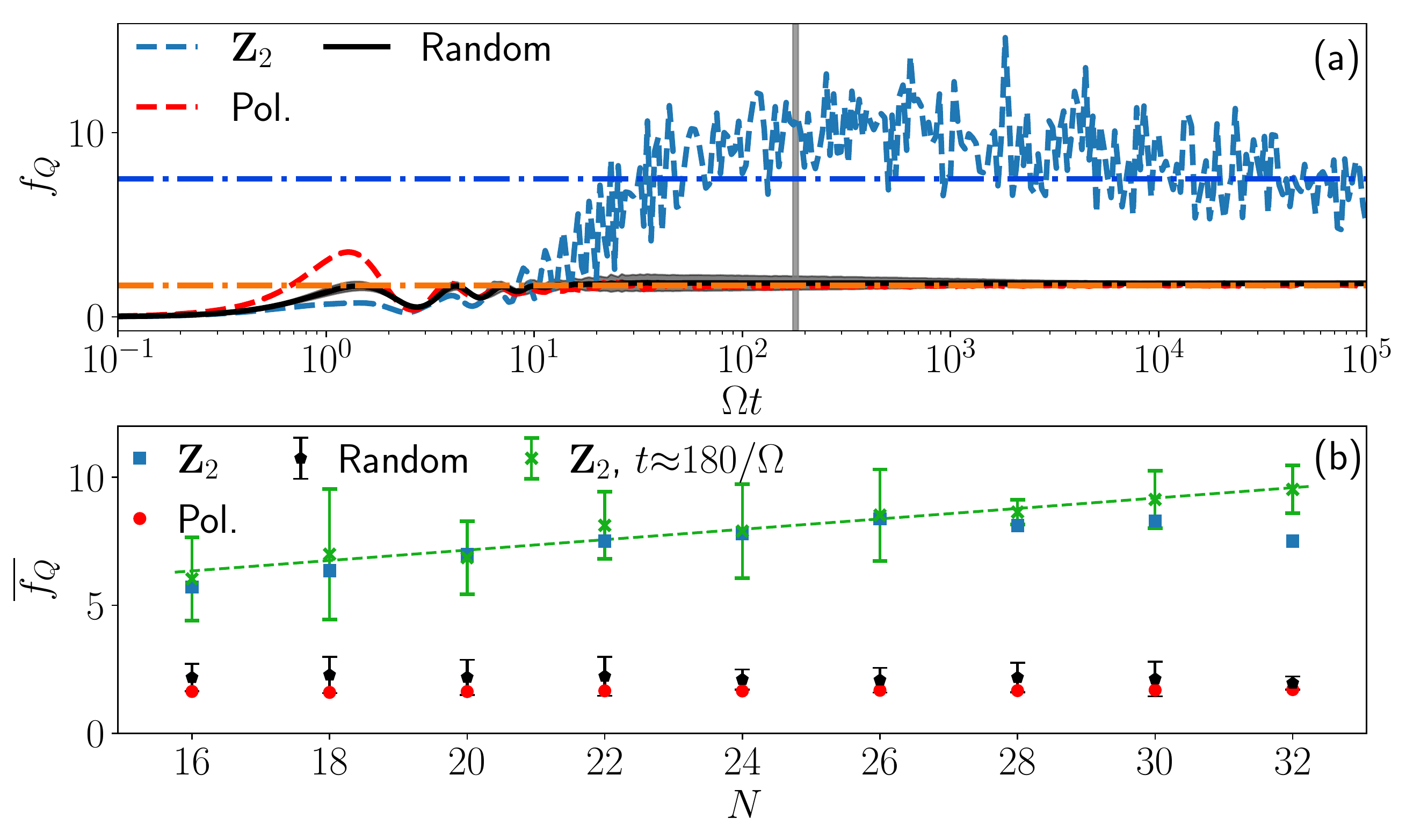}
	\caption{\small (a) Time evolution of QFI density following  quenches from various initial states. The horizontal lines show the infinite-time QFI density averages for the N\'eel state and the polarised state. 
	(b) Infinite-time QFI density averages for various initial  states as a function of system size.
	The green crosses correspond to the average over the time window indicated in grey in (a). Its width is longer than the revival period of the N\'eel state in order to average over these higher-frequency oscillations. The window is centered around $t=180/\Omega$, because at this time the maximum has been reached for all system sizes investigated but the drop at even later times has not started yet. 
	The dashed green line is a linear fit $0.20N+3.06$ to this data. In both plots the data for random states is an average over 20 samples and the error bars correspond to the standard deviation.} 
	\label{fig:PXP_QFI_dyn}
\end{figure}
While hybridisation will likely prevent any single QMBS eigenstate from having a super-extensive QFI, $F_Q{\propto}N^2$, in the asymptotic limit, such exact eigenstates cannot realistically be prepared in Rydberg atom experiments, as they lack protection from any global symmetry. Instead, we propose that extensive QFI in this model can be leveraged in practice by dynamically evolving the system to moderate times, i.e., times longer than the initial relaxation scale ${\sim}1/\Omega$, where $\Omega$ is the  Rabi frequency for the model in Eq.~\eqref{eq:H_PXP}. In Fig.~\ref{fig:PXP_QFI_dyn} we computed the evolution of QFI density when the PXP model is quenched from various initial states, contrasting the behaviour of $|\mathbb{Z}_2\rangle$ with thermalising initial states, such as the polarised state, $|{\circ}{\circ}{\circ}\cdots\rangle$, and other random product states.

The dynamics from the $|\mathbb{Z}_2\rangle$ state in Fig.~\ref{fig:PXP_QFI_dyn} clearly stands out from other thermalising initial states. Following the initial spreading, $f_Q$ undergoes a fast growth in the $|\mathbb{Z}_2\rangle$ case, reaching a broad maximum at intermediate times, $O(10^2)$.   
For all system sizes investigated (including the ones where eigenstate hybridisation is observed), the value of this maximum is extensive in system size. Larger systems can also be investigated using the symmetric subspace approximation ~\cite{turner2020correspondence}, confirming the extensive scaling within this framework~\cite{SM}. At much later times,  however, $f_Q$ starts to drop, as expected from the eigenstate plot in Fig.~\ref{fig:PXP_QFI_eigs}. The non-extensivity of the late-time value of $f_Q$ can be independently confirmed by computing the infinite-time average using the diagonal ensemble with corrections for higher moments~\cite{SM}. 

Finally, we note that in addition to the tower of $N{+}1$ scarred eigenstates considered above, the PXP model also hosts a few isolated \emph{exact} scar states near the middle of its spectrum~\cite{Lin2019}. The latter can be expressed as matrix product states and therefore have area-law entanglement. One can prove~\cite{SM} that for these states $f_Q$ is bounded by a constant when probed with the alternating magnetic field (Eq.\eqref{eq:MS}), which makes such exact PXP scars distinct from both the approximate PXP scarred eigenstates as well as the towers of exact scars obeying the restricted spectrum generating algebra in Eq.~\eqref{eq:ds}.

{\bf{Discussion}}---  In this article, we have analytically demonstrated that a large family of exact QMBS, described by Eqs.~\eqref{eq:ds}-\eqref{eq:es}, can be distinguished from bulk thermal eigenstates through the scaling of their QFI. 
We find that the long-range order of QMBS implies the super-extensive scaling of the QFI. 
This feature, together with the logarithmic scaling of the entanglement entropy, affirms the \emph{semi-classical} nature of such states, that share the same entanglement properties of asymptotic semi-classical trajectories \cite{Lerose2020Bridging}.
Moreover, our numerical study of the PXP model shows that robust signatures of superextensive QFI scaling can be expected despite the non-exact nature of QMBS in that model. We also provided evidence that this structure can be probed dynamically  by measuring the variance of an appropriate operator in current Rydberg atom experiments.
The multipartite entanglement considered here is very special and is known to have potential use for quantum enhanced metrology. Given this finding for QMBS, which are a particular example of weak-ergodicity breaking, it would be interesting to investigate if other systems in this class could show scaling of entanglement beyond bipartite correlations.

 {\bf{Acknowledgements.}}---
 We thank Alessandro Silva for useful comments on the manuscript. SP and JG thank  S. Gainsburg. J.Y.D. thanks Christopher Turner for useful comments on the diagonal ensemble. We acknowledge support by EPSRC grants EP/R020612/1 (ZP) and  EP/R513258/1 (JYD). Z.P. acknowledges support by the Leverhulme Trust Research Leadership Award RL-2019-015.  
 Statement of compliance with EPSRC policy framework on research data: This publication is theoretical work that does not require supporting research data. JG is supported by a SFI-Royal Society University Research Fellowship and acknowledges funding from European Research Council Starting Grant ODYSSEY (Grant Agreement No. 758403). SP is supported by the Simons Foundation Grant No. 454943. FP has received funding from the  European  Union’s  Horizon  2020  research  and  innovation  programme  under  the  Marie Sklodowska-Curie grant agreement No 838773.  

\bibliography{biblo.bib}

\onecolumngrid 
\newpage

\begin{center}
{\bf \large Supplementary Online Material for  
``Extensive Multipartite Entanglement from su(2) Quantum Many-Body Scars"}
\end{center}
\begin{center}
Jean-Yves Desaules$^1$, Francesca Pietracaprina$^2$, Zlatko Papi\'c$^1$, John Goold$^2$, and Silvia Pappalardi$^3$\\
\vspace*{0.1cm}
{\footnotesize
$^1$School of Physics and Astronomy, University of Leeds, Leeds LS2 9JT, United Kingdom\\
$^2$Department of Physics, Trinity College Dublin, Dublin 2, D02PN40, Ireland \\
$^3$Laboratoire de Physique de l'\'Ecole Normale Sup\'erieure, ENS, Universit\'e PSL,\\ CNRS,  Sorbonne Universit\'e, Universit\'e de Paris, F-75005 Paris, France}
\end{center}
\setcounter{subsection}{0}
\setcounter{equation}{0}
\setcounter{figure}{0}
\renewcommand{\theequation}{S\arabic{equation}}
\renewcommand{\thefigure}{S\arabic{figure}}
\renewcommand{\thesubsection}{S\arabic{subsection}}

{\footnotesize In this Supplementary Material, we provide additional analysis and background calculations to support the results in the main text. In Sec.~\ref{app:QFI_lambda} we introduce the relation between the QFI and phase-estimation theory. In Sec.~\ref{app:expVolume} we comment on exponentially decaying two-body connected correlation functions compatible with the extensive scaling of the entanglement entropy. In Sec.~\ref{app:proofQFIscars} we derive Eq.~\eqref{eq:fq_su2} of the main text. In Secs.~\ref{app:QFI_inf}-\ref{app:SU2} we provide details on the numerics of the PXP model. Finally, in Sec.~\ref{sec:mpsscars} we prove that QFI for exact scars with energies $E{=0},\pm \sqrt{2}$ in the PXP model is bounded by a constant, hence these states do not obey the extensive QFI scaling like the tower of scarred eigenstates discussed in the main text.  
}

\vspace*{1cm}

\twocolumngrid

\section{Quantum parameter estimation and Fisher information}
\label{app:QFI_lambda}
Here we provide an introduction of the quantum Fisher information through the lens of parameter estimation~\cite{Braunstein1994}, see Ref.~\cite{pezze2014quantum} for a review.
Consider a family of quantum states $\hat{\rho}_{\lambda}$ parameterised by $\lambda$. 
This parameter does not in general correspond to an observable of the system. The goal is to provide an estimation of this parameter through measurements of some observable in the system which is prepared in the state $\hat{\rho}_{\lambda}$ and optimise the inference by minimising the uncertainty. 

Measurements on quantum systems are described, in their most general form, by a positive operator valued measure (POVM) with elements $\{\hat \Pi(\zeta)\ge 0\}$ which satisfy $\int d\zeta \hat\Pi(\zeta)=\mathbb{1}$ where $\zeta$ labels possible outcomes. 
$M$ independent measurements on identical preparations of $\hat{\rho}_{\lambda}$ yields random outcomes $\zeta=\{\zeta_1,\zeta_2\dots \zeta_{M} \}$ from which a estimator function, $\lambda_{est}(\zeta_1,\zeta_2\dots \zeta_{M})$, can be constructed. 
The estimator function is a mapping from the set of random measurement outcomes to parameter space. 
The conditional probability of observing a sequence of measurement outcomes $\zeta$ given that the parameter's true value is $\lambda$ is $P(\zeta|\lambda)=\prod_{i=1}^{M}P(\zeta_{i}|\lambda)$ with $P(\zeta|\lambda)=\Tr(\hat{\rho}_{\lambda}\hat{\Pi}(\zeta))\ge 0$ and $\int d\zeta P(\zeta|\lambda)=1$ since $\int d\zeta \hat{\Pi}(\zeta)=\mathbb{1}$. 

The estimator $\lambda_{est}(\zeta)$ is a function of random variables, so it is itself a random variable with a $\lambda$ dependent mean value given by $\bar{\lambda}=\int d\zeta P(\zeta|\lambda)\lambda_{est}(\zeta)$ and variance given by $(\Delta\lambda)^2=\int d\zeta P(\zeta|\lambda)[\lambda_{est}(\zeta)-\bar{\lambda}]^2$. We will consider locally unbiased estimators such that $\bar{\lambda}$ and $\partial\bar{\lambda}/\partial\lambda=1$ so that the statistical mean of the measurement data yields the true value of the parameter. It is well known from the theory of classical parameter estimation that the variance has a fundamental lower bound known as the Cram\'er-Rao lower bound given by
\begin{equation}
\label{eq:cramer_rao_c}
    (\Delta \lambda)^2\ge(\Delta \lambda_{CR})^2=\frac{1}{M F(\lambda)} \ ,
\end{equation}
where 
\begin{equation}
    F(\lambda)=\int d\zeta P(\zeta|\lambda)\frac{\partial\log{P(\zeta|\lambda)}}{\partial\lambda} \ ,
\end{equation}
is the Fisher information and the factor of $1/M$ comes from the statistical improvement of M independent measurements. 

Is this a fundamental bound on precision for quantum parameter estimation ? The answer is no. We can do better by considering an upper-bound on the Fisher information by maximising the Fisher information over all possible measurements allowed by quantum mechanics. This leads us to the definition of the quantum Fisher information as $\mathcal{F}_{Q}(\hat{\rho}_{\lambda})=\max_{\{ \hat{\Pi} \}} F(\lambda)$ so that $\mathcal{F}_{Q}(\hat{\rho}_{\lambda})\ge F(\lambda)$ and we have the quantum Cram\'{e}r-Rao bound on the variance as 
\begin{equation}
\label{eq:quantum_cramer_rao}
    (\Delta \lambda)^2\ge(\Delta \lambda_{QCR})^2=\frac{1}{M \mathcal{F}_{Q}(\hat{\rho}_{\lambda})}.
\end{equation}
In this work we will consider states $\hat{\rho}_{\lambda}$ which are generated by some Hermitian operator $\hat{O}$ such that $\partial_{\lambda}\hat{\rho}_{\lambda}=i[\hat{\rho}_{\lambda},\hat{O}]$ and in the case an exact expression for the QFI is known to be 
\begin{equation}
\label{eq:QFI_U}
\mathcal{F}_{Q}(\hat{\rho}_{\lambda})=2\sum_{\alpha,\beta}\frac{(p_{\alpha}-p_{\beta})^2}{p_{\alpha}+p_{\beta}}|\langle \lambda_{\alpha}|\hat{O}|\lambda_{\beta}\rangle|^2
\end{equation}
where the input state is given in the diagonal form $\hat{\rho}=\sum_{\alpha}p_{\alpha}|\lambda_{\alpha}\rangle\langle\lambda_{\alpha}|$.

\section{Exponential decay of correlations and volume-law entanglement}
\label{app:expVolume}
 
The connected correlation functions evaluated on eigenstates satisfying ETH  decay exponentially at large distances [cf. Eq.~\eqref{eq:conneCh} of the main text], i.e.
\begin{equation}
    \label{eq:conneCh_SM}
    G_r(E_n) = \langle E_n | \hat o_i \hat o_{i+r} |E_n\rangle_c \sim e_r e^{-r/\xi}\ , \, r\to \infty \ .
\end{equation}
Notable exceptions include global conserved quantities, e.g., $\hat B=\sum_i \hat b_i$, which obeys $[\hat H, \hat B]=0$. In this case the variance of the conserved operator is zero in each eigenstate, $\braket{\hat B^2}_c=0$, imposing a sum rule that leads to  $\braket{\hat b_i \hat b_{i+r}}\sim -\frac{1}{2(N-1)}$, see e.g. Ref.~\onlinecite{Pal2010MB}. 
The scaling in Eq.~\eqref{eq:conneCh_SM}, despite being unambiguous classically, might look counterintuitive with respect to our knowledge about quantum chaotic eigenstates. These are well known to be highly entangled quantum states, characterized by a volume law scaling of the entanglement entropy \cite{Deutsch2013Microscopic, Vidmar2017Entanglement, Murthy2019Structure}.  In other words, they can not be represented as matrix product states (MPS) of finite bond dimension. This might seem to contradict Eq.~\eqref{eq:conneCh_SM}, which states the absence of long-range correlations between local operators. 

Few remarks are in order.
First of all, the connected correlation functions $G_r(E_n)$ of local (small) operators encode only the \emph{local} information of the two-point reduced density matrix (on sites $i$ and $i+n$). Hence, their scaling makes no predictions on the non-local structure encoded in the entanglement entropy of the --- usually large --- region $A$. Such entanglement entropy $S_A$ is defined as the von Neumann entropy of the reduced density matrix $\hat \rho_A$ and to study its scaling with the system size one typically lets the region $A$ scale with the volume of the system. Therefore, 
	to falsify the volume law scaling, one would need exponential decay of correlations of non-local (large) operators with support over $A\propto \text{Vol}$. This statement is much stronger than Eq.~\eqref{eq:conneCh_SM} and it is usually wrong. Interestingly, this occurs for many-body-localized (MBL) eigenstates \cite{Abanin2019Colloquium}. 
A final comment concerns the MPS representation of the eigenstates. While one could always describe the two-point reduced density matrix (on sites $i$ and $i+n$) as an MPS of finite bond dimension, the MPS representation of the full eigenstate constitutes a \emph{global} description and, for what argued above, the site in the middle of the system should carry the non-local correlations for size $N/2$.\\

This comment highlights in what sense chaotic ETH eigenstates are different from thermal density matrices $\hat \rho_{\text{Gibbs}}$. They have the same correlations as local operators, but when it comes to non-local ones they can be very different. Indeed, the mutual information of $\hat \rho_{\text{Gibbs}}$ is area law and this state can be written efficiently as MPO. On the other hand, chaotic eigenstates are more complicated, hosting volume law entanglement and non-local correlations.

\section{Genuine multipartite entanglement of exact scars}
\label{app:proofQFIscars}

Here we provide a detailed derivation of Eq.~\eqref{eq:fq_su2} of the main text. 
The operators $\hat J^{\pm}$ and $\hat J^z$ obey the su(2) algebra on the subspace $W$. Then one can define a generalisation of the collective spin, i.e.
 \begin{align}
 	\hat{\bold J}^2  & = (\hat J^x)^2 + (\hat J^y)^2 + (\hat J^z)^2 
 	= (\hat J^z)^2  
 	+ \frac 12 \left ( \hat J^+ \hat J^- +\hat  J^-\hat  J^+ \right ) \nonumber 
 	\\ 
 	& =_w\hat J^z \hat (J^z + 1) +\hat  J^-  \hat J^+ =_w \hat  J^z (\hat J^z - 1) + \hat J^+  \hat J^- \ .
 \end{align}
 Since $[\bold J^2 , J^{z, \pm}]=_w 0$, the scars $\ket{\mathcal S_n}$ are characterized by a fixed eigenvalue of $\bold J^2$ that can be determined directly from the ground state as
 \begin{align}
 	\hat {\bold J}^2 \ket{\mathcal S_0} 
 	& = \left[  \hat J^z (\hat J^z - 1) +\hat  J^+  \hat J^- \right ]\ket{\mathcal S_0}
 	= \frac{E_0}{\omega} \left ( \frac{E_0}{\omega}  - 1 \right) \ket{\mathcal S_0} \ ,
 \end{align}
 where we have used $J^- \ket{\mathcal S_0} = 0$ and the definition of $\hat J^z$ in Eq.~\eqref{eq:SU2} of the main text. This is equivalent to a total spin $S = -E_0/\omega$. We can now compute 
 \begin{align}
    \label{eq:StarIAde}
 	\bra{\mathcal S_n} \hat J^+ \hat J^- \ket{\mathcal S_n} 
 	& = \bra{\mathcal S_n} \hat {\bold J}^2  -   \hat J^z (\hat J^z - 1)\ket{\mathcal S_n} \\
 	& = \frac{E_0}{\omega} \left ( \frac{E_0}{\omega}  - 1 \right) - \frac{E_n}{\omega} \left ( \frac{E_n}{\omega}  - 1 \right)\\
 	\label{eq:eiVal}
 	& = - 2\frac {E_0}{\omega} n - n^2 + n \ .
 \end{align}
 If we now use the extensivity of the ground state energy $E_0 = -\epsilon_0 N$ and divide everything by $N^2$ we directly obtain Eq.~\eqref{eq:lro} in the main text.   Eq.~\eqref{eq:StarIAde} gives back exactly Eq.~\eqref{eq:es} of Schecter and Iadecola \cite{Iadecola2019_2} with $\omega= 2 h$ and $\epsilon_0 = h$.

\section{Infinite-time average of the quantum Fisher information}
\label{app:QFI_inf}
To get the value of the quantum Fisher information in the long-time limit after a quench, one way is to compute it from the diagonal ensemble. 
Indeed, assuming a non-degenerate spectrum it holds that for an operator $\hat{O}$
\begin{equation}
\overline{\langle \hat{O}\rangle}_{\infty}\hspace{-0.1cm}=\hspace{-0.1cm}\lim\limits_{T\rightarrow \infty}\frac{1}{T}\int dt \braket{\psi(t)| \hat O|\psi(t)}{=}\hspace{-0.1cm}\sum_n |c_n|^2O_{n,n},
\end{equation}
where $O_{i,j}=\braket{E_i|\hat O|E_j}$ and $c_n=\braket{E_n|\psi}$.

In the case of the PXP model we also need to consider the large number of ``zero modes"~\cite{TurnerPRB, Iadecola2018} and treat them separately, leading to 
\begin{equation}\label{eq:diagE}
\begin{aligned}
\overline{\langle \hat O\rangle}_{\infty}&=\lim\limits_{T\rightarrow \infty}\frac{1}{T}\int dt \braket{\psi(t)|\hat{O}|\psi(t)}\\
&=\sum_{\substack{n\ \text{s.t.}\\ E_n\neq 0}} |c_n|^2O_{n,n}+\sum_{\substack{n,m\ \text{s.t.}\\ E_n=E_m=0}} c_n^\star c_m O_{n,m}.
\end{aligned}
\end{equation}

To simplify the notation, we introduce the diagonal ensemble density matrix $\hat{\rho}$, such that $\rho_{m,n}=c_m^\star c_n \delta(E_m-E_n)$.
In that case we simply have $\overline{\langle \hat{O}\rangle}_{\infty}=\Tr{\hat \rho \hat{O}}$.
For the PXP model this allows to compute the infinite-time average of the staggered magnetisation $\hat{M}_S$ and of its square $\hat{M}_S^2$.
However in order to get the QFI we also need the infinite-time average of $\braket{\hat O}^2$.
Explicitly computing this quantity leads to the more complicated expression
\begin{equation}
\begin{aligned}
&\lim\limits_{T\rightarrow \infty}\frac{1}{T}\int dt \, \langle \hat{O}\rangle^2 \\
{=}&\hspace{-0.25cm}\sum_{a,b,c,d}c_a^\star c_b c_c^\star c_d O_{a,b}O_{c,d} \lim\limits_{T\rightarrow \infty}\frac{1}{T}\hspace{-0.1cm}\int \hspace{-0.15cm}dt \, {\rm e}^{ i(E_a-E_b+E_c-E_d)}\\
{=}&\hspace{-0.25cm}\sum_{a,b,c,d}c_a^\star c_b c_c^\star c_d O_{a,b}O_{c,d}\delta(E_a-E_b+E_c-E_d),
\end{aligned}
\end{equation}
with a Dirac delta now holding 4 energies instead of 2.
Fortunately, as the number of combinations that satisfy it is relatively limited, this is still tractable numerically.

The simplest case is $E_a=E_b$, $E_c=E_d$, which can be rewritten as 
\begin{equation}
\begin{aligned}
\sum_{a,d}|c_a|^2|c_c|^2 O_{a,a}O_{c,c}&{=}\hspace{-0.12cm}\left(\sum_a  |c_a|^2 O_{a,a}\right)\hspace{-0.15cm}\left(\sum_c  |c_c|^2 O_{c,c}\right)\\
&=\left( \Tr{\hat{\rho} \hat{O}} \right)^2{=}\left(\overline{\langle \hat{O}\rangle}_{\infty}\right)^2
\end{aligned}
\end{equation}
Another possibility to satisfy the Dirac delta is to have $E_a=E_d$ and $E_b=E_c$. 
In that case we have
\begin{equation}
\begin{aligned}
&\sum_{a,b}|c_a|^2|c_b|^2 O_{a,b}O_{b,a} \\
{=}&\sum_{a,b}\braket{\psi|E_a}\braket{E_a|O|E_b}\braket{E_b|\psi}\braket{\psi|E_a}\braket{E_b|\hat O|E_a}\braket{E_a|\psi}\\
{=}&\Tr{\hat \rho \hat O \hat \rho \hat O}.
\end{aligned}
\end{equation}

Finally, as in the PXP model the spectrum is symmetric around $E{=}0$ the last possibility is $E_a{=}-E_c$ and $E_b{=}-E_d$.
However for $\hat{M}_S$ in the momentum sectors $k{=}0$ and $k{=}\pi$ (where the N\'eel state has support) we can take advantage of the eigenstates, the operator, and the initial state  being real.
Indeed, as the Hamiltonian projected to one of these sectors is real and symmetric the eigenvectors themselves can be chosen to be real.
In that case $c_a=c_a^\star$ (as all initial states are real in the Fock basis) and $(\hat{M}_S)_{a,b}=(\hat{M}_S)_{b,a}$ since the matrix is symmetric (as it is both real and Hermitian).
We can then rewrite 
\begin{equation}
\begin{aligned}
&\sum_{a,b}c_a^\star c_b c_a^\star c_b O_{a,b}O_{a,b}\\
=&\sum_{a,b}c_a^2 c_b^2 O_{a,b}O_{b,a}\sum_{a,b}|c_a|^2|c_b|^2 O_{a,b}O_{b,a} \\
=&\Tr{\hat{\rho} \hat{O} \hat{\rho} \hat{O}}
\end{aligned}
\end{equation}
which is the same as the contribution with $E_a{=}E_d$ and $E_b{=}E_c$.
Summing all these contribution together we can compute the long-time average of the QFI in the PXP model in the sectors $k=0$ and $k=\pi$ for a real-valued operator $\hat{O}$ as
\begin{equation}
\begin{aligned}
&\lim\limits_{T\rightarrow \infty}\frac{1}{T}\int dt \, 4\left(\braket{\psi(t)| \hat{O}^2 |\psi(t)}-\braket{\psi(t)| \hat{O} |\psi(t)}^2\right)\\
=&4\, \Tr{\hat{\rho} \hat{O}^2}-4\left(\Tr{\hat{\rho} \hat{O}}\right)^2-8\, \Tr{\hat{\rho} \hat{O} \hat{\rho} \hat{O}}.
\end{aligned}
\end{equation}
 
\section{Long-range order in scarred PXP eigenstates}
\label{app:LR_PXP}
In this section we look at the connected correlation function for the eigenstates of the PXP model.
As the operator used to probe the QFI of eigenstates in this model is the staggered magnetisation $\hat{M}_S=\frac{1}{2}\sum_j (-1)^j \hat Z_j$, we investigate 
 \begin{equation} \label{eq:corre}
 G_r(E_n)=\braket{E_n|\hat Z_j \hat Z_{j+r}|E_n}-\braket{E_n|\hat Z_j|E_n}\braket{E_n|\hat Z_{j+r}|E_n}.
 \end{equation}
In the approximate algebra proposed in \cite{Iadecola2019}, the scarred states are eigenstates of $\hat{J}^z$ and $\hat{M}_S$ is equivalent to $\hat{J}^x$.
Hence both raising and lowering operators depend on $\hat{M}_S$ and the correlation function in Eq.~\eqref{eq:corre} is proportional to a term of the one-excitation reduced density matrix $\braket{ \hat J^+ \hat J^-}$.

\begin{figure}[t]
	\centering
	\includegraphics[width=\linewidth]{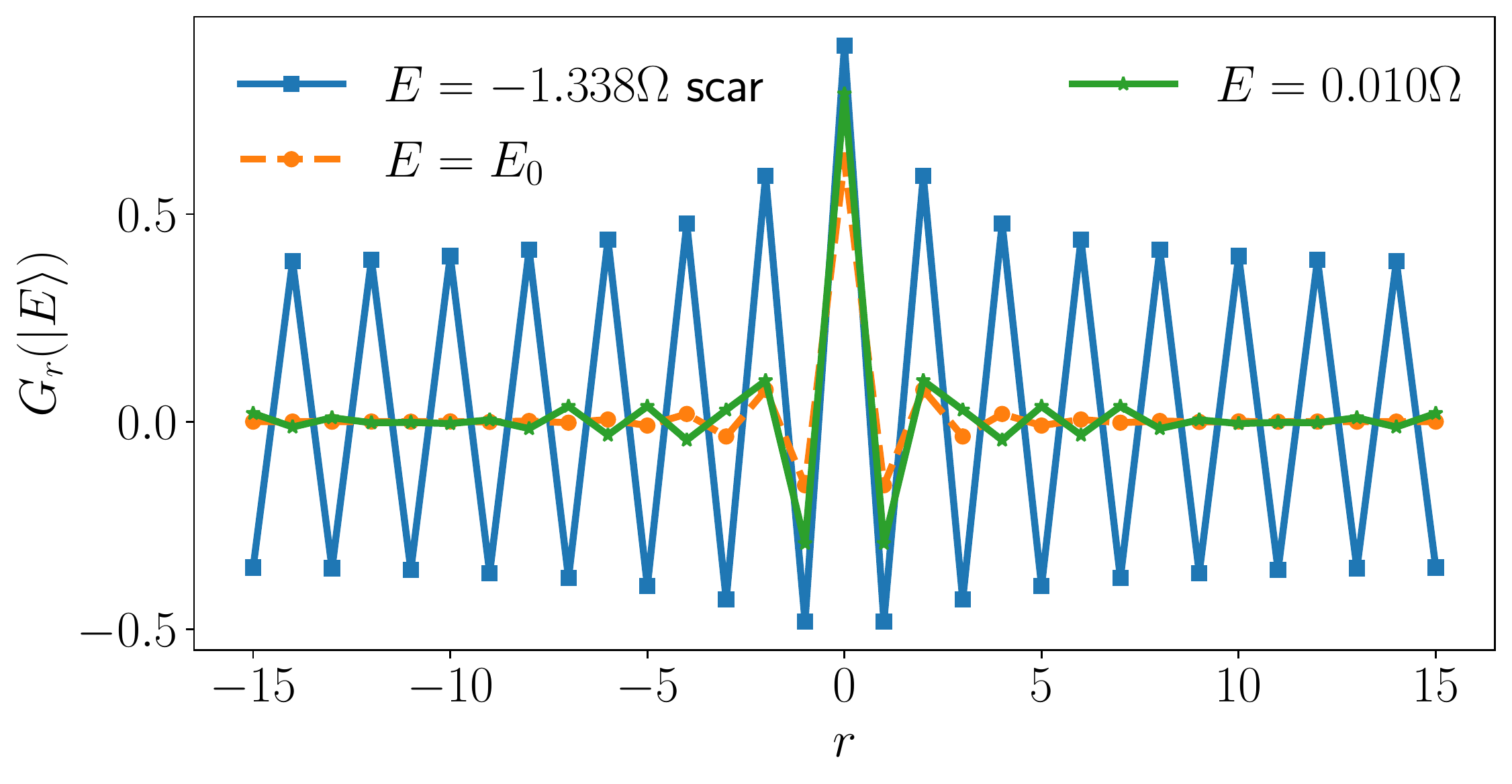}
	\caption{Connected correlation function \eqref{eq:corre} for eigenstates of the PXP model with $N=30$. The scarred eigenstates near the middle of the spectrum ($E=\, -1.338$) exhibit long-range order in contrast  with the thermal states or states at the edge of the spectrum.}
	\label{fig:PXP_ZZ_corr}
\end{figure}

From Fig.~\ref{fig:PXP_ZZ_corr} it is clear that 1) correlations evaluated on thermal eigenstates decay exponentially with $r$ [cf. Eq.~\eqref{eq:conneCh}] 2) the decay of correlations for scars is much slower for scarred eigenstates than for thermal ones. 
However for large enough system the connected correlation function is not well described by an exponential decay to a non-zero value.
Indeed, $G_r$ still decays for large values of $r$, albeit slowly.

\section{Extensive Fisher information in large PXP eigenstates within a symmetric subspace approximation}
\label{app:Sym_PXP}
Unlike a few simple eigenstates with energy $E{=}0$~\cite{Lin2019}, most  scarred PXP eigenstates  cannot be constructed exactly.  Therefore,  results on eigenstates for this model are limited to system sizes for which exact diagonalisation is possible. 
However, approximation schemes have been developed for this model, allowing to access much larger systems. 
In particular, the symmetric subspace approximation~\cite{turner2020correspondence} allows to accurately approximate scarred PXP eigenstates while it maintains a direct connection with the semiclassical limit of the PXP model described by the Time-Dependent Variational Principle (TDVP)~\cite{wenwei18TDVPscar}.

The symmetric subspace approximation is based on the observation that scarred eigenstates are approximately invariant under permutations that do not exchange odd and even sites in the chain.  This is not a symmetry of the full PXP model, as the latter has strictly local interactions. Note that because of the Rydberg blockade condition, not all of the permutations are allowed, and the way to enforce this approximate symmetry is to group all states into equivalence classes.

Let us denote by $n_1$ the number of excitations present on the even sublattice (the sublattice encompassing all even sites) and by $n_2$ their number on the odd sublattice.
We can then group all states with the same values of $n_1$ and $n_2$ into an equivalence class, and define the state $\ket{n_1,n_2}$ as the symmetric superposition of all states in that class.
From there, the symmetric subspace $\mathcal{K}$ is defined as the span of all possible $\ket{n_1,n_2}$.
It is easy to see that any state in $\mathcal{K}$  will be invariant under a permutation that does not mix even and odd states and does not violate the Rydberg blockade.
It is also straightforward to show that the dimension of $\mathcal{K}$ only grows quadratically with $N$.
The projection  of the PXP Hamiltonian into this subspace leads to approximate eigenstates referred to as ``quasimodes", denoted as $\ket{E_\mathcal{K}}$. Specifically, for each scarred PXP eigenstate there is a corresponding quasimode with a high overlap with it~\cite{turner2020correspondence}.
We will denote the approximate eigenstates corresponding to the scarred ones as "top-band quasimodes".
The sublattice permutation symmetry then protects these states from the exponential density of states, and no hybridisation is visible even for a few hundred sites. 

\begin{figure}[t]
	\centering
	\includegraphics[width=\linewidth]{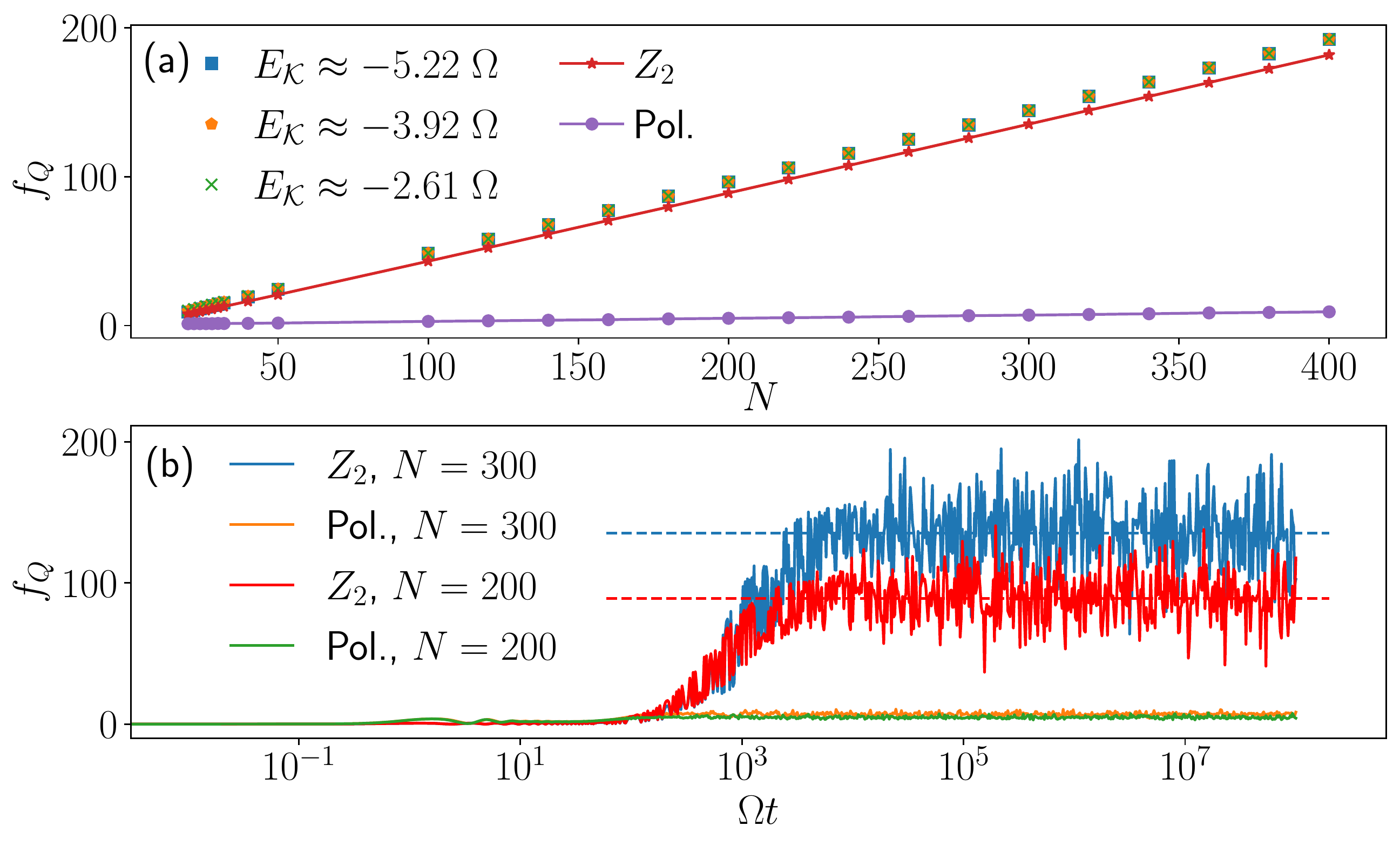}
	\caption{\small (a) QFI density of top-band quasimodes near the middle of the spectrum and the infinite-time average for the N\'eel and polarised states. Both are clearly extensive in $N$. (b) Time evolution of the QFI density in quenches from the N\'eel and polarised states for two different system sizes. The horizontal dashed lines are the infinite-time averages for the N\'eel state. As they match the value of the plateau we expect no drop of $f_Q$ even at very late times. }
	\label{fig:PXP_perm_QFI}
\end{figure}

Performing exact diagonalisation in $\mathcal{K}$  shows that the quantum Fisher information of the top-band quasimodes is extensive in system size, as can be seen in Fig.~\ref{fig:PXP_perm_QFI}. 
Accordingly, when quenching from the N\'eel state the QFI reaches a plateau (with a value also super-extensive in $N$) and does not drop down afterwards. 
This is confirmed by the computation of the infinite time average that matches the value of the plateau. 
Note that the super-extensive scaling of the QFI after a quenched dynamics occurs also for fully-connected semi-classical spin systems \cite{Lerose2020Bridging}. 

\begin{figure}[t]
	\centering
	\includegraphics[width=\linewidth]{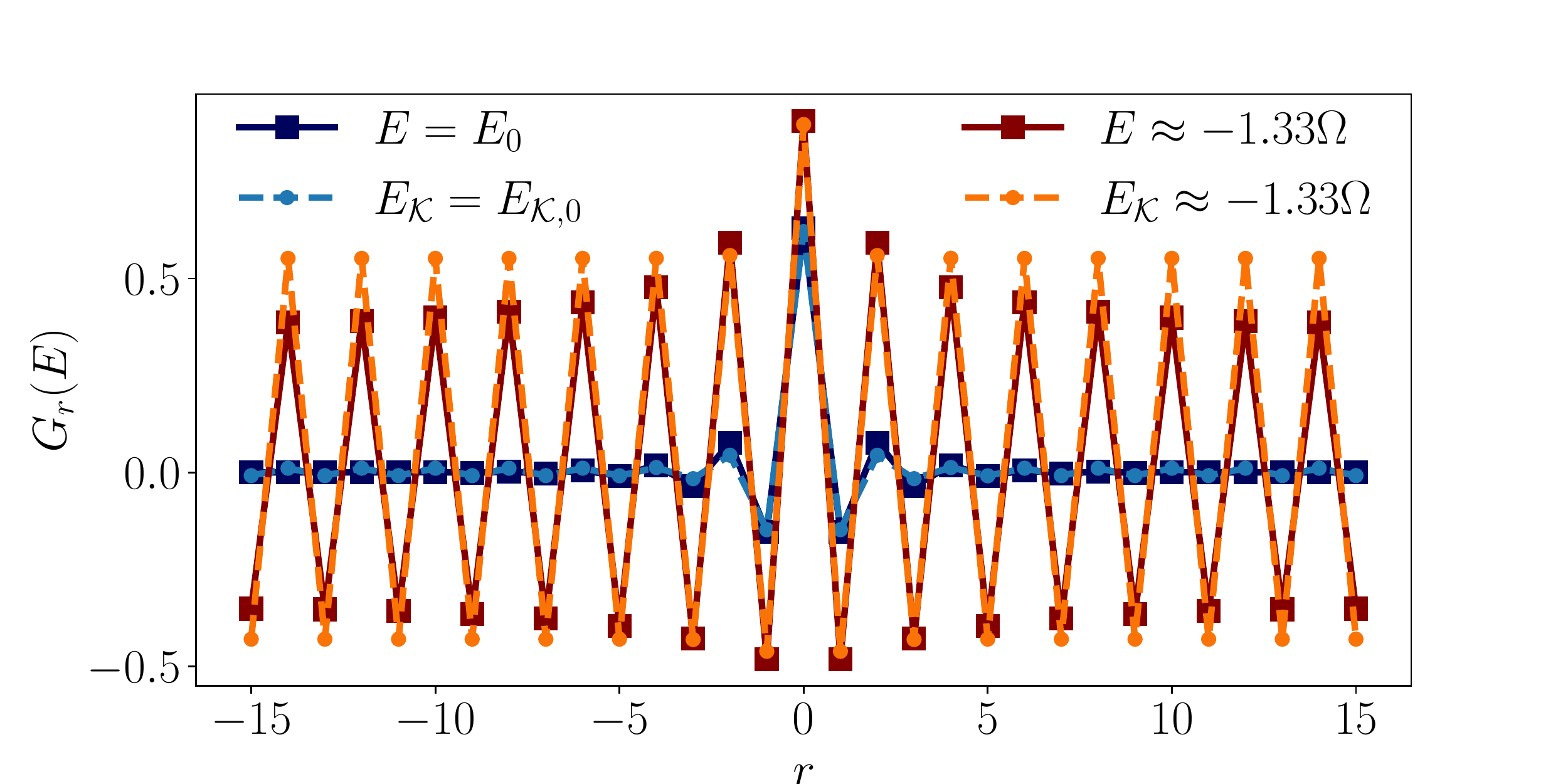}
	\caption{\small $Z_jZ_{j+r}$ correlation function for  the eigenstates of the PXP model and their corresponding quasimodes for $N=30$ spins.}
	\label{fig:PXP_corr_comp}
\end{figure}

\begin{figure}[b]
	\centering
	\includegraphics[width=\linewidth]{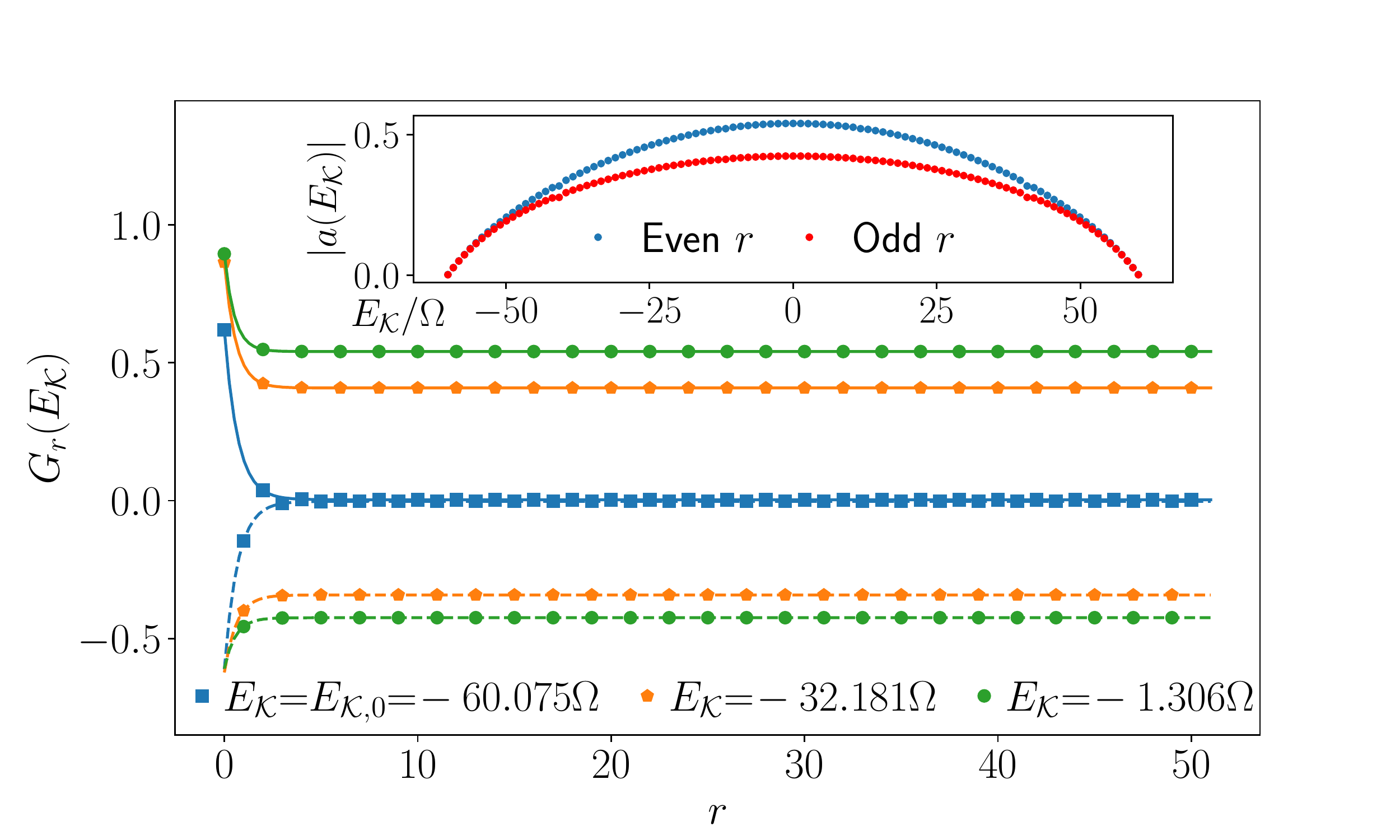}
	\caption{\small $Z_jZ_{j+r}$ correlation function for three top-band quasimodes with $N=100$. The inset shows the constant part of $G_r$ obtained by fitting the numerical results with Eq.~\eqref{eq:Gr_sym}. Odd and even values of $r$ have been fitted separately. The non-zero value of $a$ show that the top-band quasimodes exhibit long-range order even for very large systems.}
	\label{fig:PXP_corr_perm}
\end{figure}

Comparing the connected correlation function also shows a different behaviour in $\mathcal{K}$ (Fig.~\ref{fig:PXP_corr_comp}). 
Where the exact scarred eigenstates show a slow decay of $G_r$, their quasimodes counterparts very clearly follow an exponential decay to a non-zero value as
\begin{equation}\label{eq:Gr_sym}
    G_r=a+b{\rm e}^{-r}.
\end{equation}
This is also true for much larger systems and there we can fit $G_r$ using \eqref{eq:Gr_sym} to see that $a$ becomes a smooth function of the energy density (Fig.~\ref{fig:PXP_corr_perm}). 

Overall, the top-band quasimodes show that same phenomenology as \emph{exact} quantum many-body scars.
As the true scarred eigenstates have a high overlap on them, they also show anomalous values for the QFI and for correlation functions. This behaviour, reminiscent of exact scars, is likely due to the fact that hybridisation with the thermal bulk is explicitly suppressed in the construction of the quasimodes.

\section{Emergent su(2) algebra in the PXP scarred subspace}
\label{app:SU2}

The PXP model exhibits long-lived periodic revivals when initialised in the N\'eel $|\mathbb{Z}_2\rangle$ state,  despite its large energy density (formally corresponding to an infinite temperature).  The revivals are accompanied by a
generally linear growth of the  bipartite entanglement entropy, which is slower compared to other thermalising initial states. It has been understood~\cite{turner2018weak, Choi2018, Bull2020} that such dynamics arise due to
the existence of a band of non-thermal (``scarred")  eigenstates that are approximately equally spaced in energy, and have large overlaps with $\ket{\mathbb{Z}_2}$ state. This gives rise to an emergent su(2) algebraic structure of such eigenstates, as we summarise below.

The scarred PXP eigenstates can be approximately constructed using an analytical framework dubbed the forward scattering approximation (FSA)~\cite{turner2018weak, TurnerPRB}. The FSA relies on decomposing the PXP Hamiltonian into a ``raising'' and ``lowering'' part, $\hat{H}_\mathrm{PXP}=\hat{H}^+_\mathrm{PXP}+\hat{H}^-_\mathrm{PXP}$, with 
\begin{equation}
\hat{H}^\pm_\mathrm{PXP} =  \sum_{i\in \text{even}}  \hat{\mathcal{C}} \hat{\sigma}_i^{\pm} \hat{\mathcal{C}}+  \sum_{i\in \text{odd}}  \hat{\mathcal{C}} \hat{\sigma}_i^{\mp} \hat{\mathcal{C}},    
\end{equation}
where $\hat{\mathcal{C}} = \prod_i[1- (1+\hat{\sigma}^z_i)(1+\hat{\sigma}^z_{i+1})/4]$ is a global projector enforcing the Rydberg blockade condition. These operators are defined to simultaneously excite an atom on an even sublattice of the chain or deexcite an atom on the odd sublattice (and vice versa). 

Using the raising operator, one further defines its Krylov subspace $\mathcal{M}$ obtained by repeated action on the $|\mathbb{Z}_2\rangle$ state:
\begin{equation}
    \mathcal{M} = \mathrm{span}\{ |k\rangle \equiv \beta_k (\hat{H}^+_\mathrm{PXP})^k  |\mathbb{Z}_2\rangle \},
\end{equation}
with $k\in\{0, 1, 2, \dots, N\}$ and $\beta_k$ the normalisation of each vector. Note that the subspace $\mathcal{M}$ is only of dimension $N+1$ because the  vector $|k\rangle$ has a Hamming distance $k$  relative to $|\mathbb{Z}_2\rangle$ state. This is because only the raising part of the Hamiltonian  is used in defining $\mathcal{M}$, hence the name ``forward" scattering approximation.  It has been shown that scarred PXP eigenstates are predominantly supported by these FSA vectors spanning the subspace $\mathcal{M}$~\cite{turner2018weak}.

The accuracy of the FSA, and therefore the stability of revivals, relies on the dynamics of $\ket{\mathbb{Z}_2}$ generated by $\hat{H}^\pm_\mathrm{PXP}$ being (nearly) closed in the subspace $\mathcal{M}$. 
This condition would be exactly achieved if the vectors $|k\rangle$ were eigenstates of the operator
\begin{equation}\label{eq:PXP_Hz}
 \hat{H}^z \equiv[\hat{H}_0^+,\hat{H}_0^-],   
\end{equation}
but this is not  satisfied for $2\leq k \leq N-2$.
In Ref.~\cite{Choi2018} (see also Ref.~\cite{Khemani2019}) it was found that the FSA error can be suppressed by many orders of magnitude if the PXP Hamiltonian is deformed, $\hat{H}_\mathrm{PXP} \mapsto \hat{H}_\mathrm{PXP} + \delta \hat{H}_R$, by adding to it the term
\begin{equation}\label{hk}
\delta \hat{H}_R = -\sum_i \sum_{d =2}^R  h_d \; \hat{\mathcal{C}} \hat{\sigma}^x_i  \hat{\mathcal{C}} \left( \hat{\sigma}^z_{i-d} + \hat{\sigma}^z_{i+d} \right).
\end{equation}
This term introduces additional interactions between pairs of spins separated by a distance $d$, with exponentially decaying strengths 
\begin{align}
\label{eqn:ansatz}
h_d =  h_0 \left( \phi^{(d-1)} - \phi^{-(d-1)}\right)^{-2},
\end{align}
where $\phi = \left(1+\sqrt{5}\right)/2$ is the golden ratio, and $h_0 \approx 0.051$. is a (numerically determined) strength of the deformation. Evidently, the deformation is weak compared to the energy scale of the PXP Hamiltonian since $h_0 \ll 1$. 

 \begin{figure}[tbh]
	\centering
	\includegraphics[width=\linewidth]{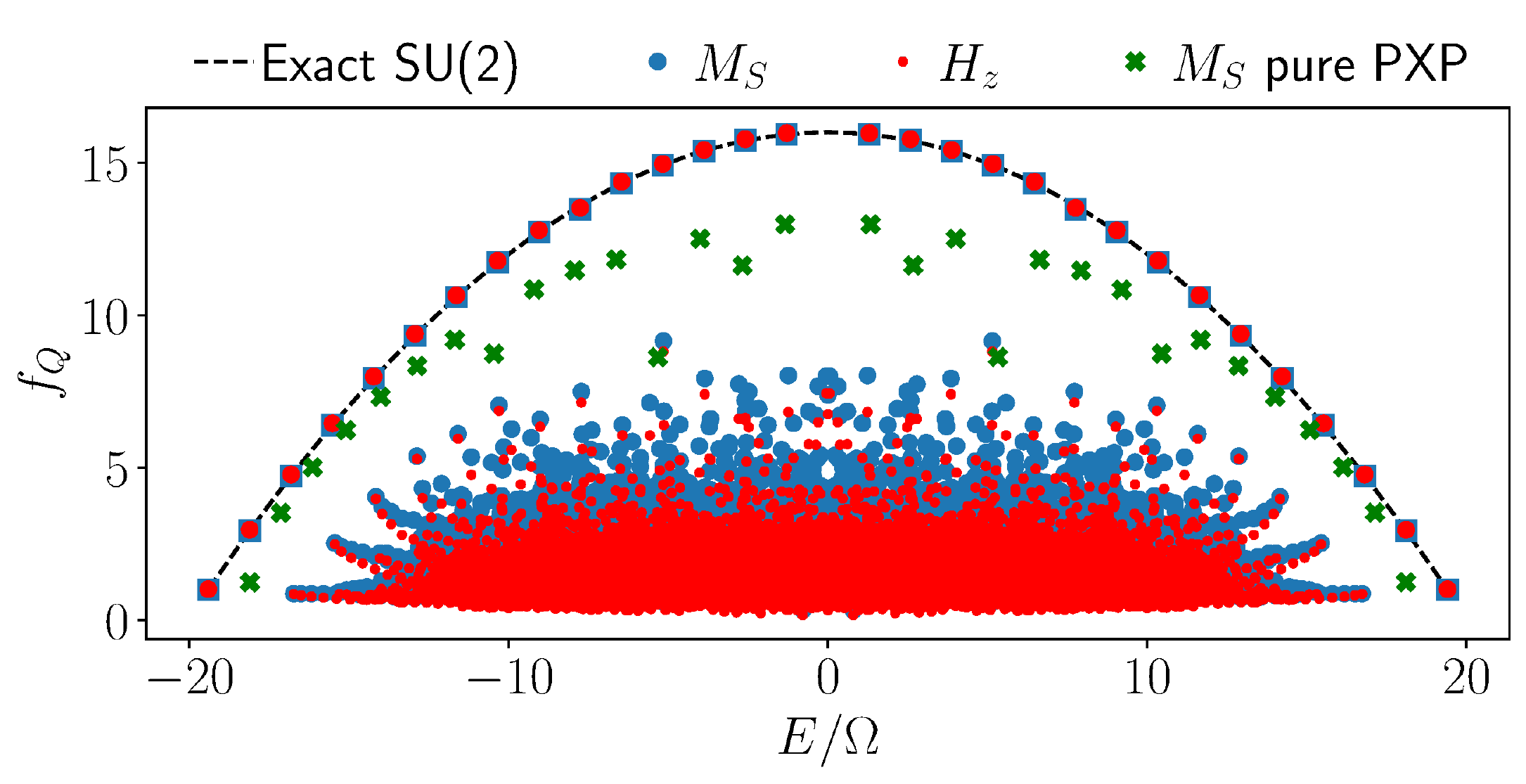}
	\caption{\small QFI density of the PXP eigenstates with and without perturbation probed with the operators $H_z$ [see Eq.~\eqref{eq:PXP_Hz}] and $M_S$ for $N=30$. For both operators, the QFI density of the scarred eigenstates of the perturbed PXP model shows good agreement with the expectations based on the exact su(2) algebra. In the perturbed case, no hybridisation is visible, unlike in the pure PXP model.}
	\label{fig:PXP_SU2_QFI}
\end{figure}
    
The model $\hat{H}_\mathrm{PXP} + \delta \hat{H}_R$ exhibits almost perfect revivals from $|\mathbb{Z}_2\rangle$ initial state~\cite{Choi2018}, suggesting that deformed $\hat{H}_\mathrm{PXP}^\pm$, i.e., after an appropriate  replacement 
\begin{equation}
 \hat{\sigma}_i^\pm \mapsto \hat{\sigma}_i^\pm \left(1+\sum_d h_d (\hat{\sigma}_{i+d}^z + \hat{\sigma}_{i-d}^z)\right),   
\end{equation}
form the su(2) ladder operators. 
Indeed, it has been numerically shown that 
\begin{align}
\label{eqn:su2}
\hat{P}_\mathcal{M} [\hat{H}^z, \hat{H}^\pm] \hat{P}_\mathcal{M} \approx  \pm \Delta \hat{P}_\mathcal{M} \hat{H}^\pm \hat{P}_\mathcal{M},
\end{align}
where $\hat{P}_\mathcal{M} = \sum_k \ket{k}\bra{k}$   is the projector onto the subspace, and $\Delta$ is a constant. As $|0\rangle = |\mathbb{Z}_2\rangle$ is an eigenstate of $\hat{H}^z$, $|k\rangle$ are also approximate eigenvectors of $\hat{H}^z$ with harmonically spaced eigenvalues $H_k^z = \langle k | \hat{H}^z | k \rangle$ so that $\Delta = H^z_{k+1} - H^z_k$. Thus, upon a suitable rescaling, the operator $\hat{H}^z$ plays the role of $\hat{J}^z$ in the su(2) algebra, and $\hat{H}^\pm$ play the role of spin-raising and lowering operators within $\mathcal{M}$. As the dimensionality of the subspace $\mathcal{M}$ is $N+1$, this implies that the operators form a spin $s = N/2$ representation of the su(2) algebra, with $|\mathbb{Z}_2\rangle$ and its translated version $|{\mathbb{Z}'_2}\rangle$ being the lowest and highest weight states,  respectively.  Thus, the virtually perfect oscillatory dynamics of $|\mathbb{Z}_2\rangle$ can be understood as a large spin $(s= N/2)$ pointing initially in an emergent ``$z$-direction'', undergoing a coherent Rabi oscillation under the Hamiltonian $\hat{H} = \hat{H}^+ + \hat{H}^-$, which is akin to the $\hat{J}^x$ operator, with period $\tau  = 2\pi/\sqrt{2\Delta}$. The emergence of this su(2) structure within $\mathcal{M}$ is nontrivial, since the PXP Hamiltonian by itself does not have any rotational symmetry.   

Fig.~\ref{fig:PXP_SU2_QFI} shows the QFI density of the PXP eigenstates with and without perturbation, probed with the operator $\hat{H}_z$ in Eq.~\eqref{eq:PXP_Hz} as well as $\hat{M}_S$, for a system of $N=30$ spins. 
We note that $\hat{M}_s$ is related to the proposed approximate su(2) algebra in Ref.~\cite{Iadecola2019}, which is not identical to that of Ref.~\onlinecite{Choi2018}. Nevertheless, it is conjectured that a perturbation that makes one of them exact will also improve the other algebra, and that at that point the two of them differ only by a change of basis~\cite{Iadecola2019}.
For both operators, the QFI of the scarred eigenstates of the perturbed PXP model is in agreement with expectations based on the perfect su(2) algebra. The enhanced proximity to the exact su(2) is also seen to strongly suppress hybridisation, compared to the pure PXP model.

\section{Exact scars in the PXP model}\label{sec:mpsscars}
In the PXP model, the $N{+}1$ scarred eigenstates discussed in the main text do not have a known analytic form. However, a few \emph{exact} eigenstates with low entanglement entropy near the middle of the spectrum admit an exact expression using matrix product states (MPS) for chains of even size~\cite{Lin2019}. In this section we point out that these exact PXP scarred eigenstates are distinct from the ones studied in the main text, in that their QFI is bounded by a constant. 

The states considered in Ref.~\cite{Lin2019} are defined using the bond-dimension 2 and 3 matrices 
\begin{eqnarray}
B^0 &=&
\begin{pmatrix}
1 & 0 & 0 \\
0 & 1 & 0
\end{pmatrix} ~, ~~~~~
B^1 = \sqrt{2}
\begin{pmatrix}
0 & 0 & 0 \\
1 & 0 & 1
\end{pmatrix} ~, \\
C^0 &=&
\begin{pmatrix}
0 & -1 \\
1 & 0 \\
0 & 0
\end{pmatrix} ~, ~~~~~
C^1 = \sqrt{2}
\begin{pmatrix}
1 & 0 \\
0 & 0 \\
-1 & 0
\end{pmatrix} ~.
\end{eqnarray}
For periodic boundary condition (PBC), it was shown that the state
\begin{eqnarray} \label{eqn:exactG}
\ket{\Phi_1} &=& \sum_{\{\sigma \}} \Tr{B^{\sigma_1} C^{\sigma_2} \dots B^{\sigma_{L-1}} C^{\sigma_L}} \ket{\sigma_1 \dots \sigma_L } ~. 
\end{eqnarray} 
and its translated version $\ket{\Phi_2}$ (for which $B$ and $C$ are swapped) are eigenstates of the PXP model (Eq.~\eqref{eq:H_PXP} in the main text) with energy $E{=}0$.
Similarly, for open boundary condition the states
\begin{equation}
\ket{\Gamma_{\alpha, \beta}} = \sum_{\{\sigma\}} v_\alpha^T B^{\sigma_1} C^{\sigma_2} \dots B^{\sigma_{L-1}} C^{\sigma_L} v_\beta \ket{\sigma_1 \dots \sigma_L} ~,
\label{eqn:GammaOBC}
\end{equation}
are eigenstates, where $\alpha,\ \beta=1,2$ and the corresponding boundary vectors are $v_1 = (1, 1)^T$ and $v_2 = (1, -1)^T$.
Their energies are $E_{1,1}{=}E_{1,2}=0$, $E_{1,2}{=}\sqrt{2}$ and $E_{2,1}{=}\; -\sqrt{2}$.

We probe the QFI density of these states using the staggered magnetisation (Eq.~\eqref{eq:MS} in the main text).
The computation of QFI can be done in a straightforward manner by making use of the blocked representation introduced in Ref.~\cite{Lin2019}, and by working in the diagonal basis of the transfer matrix.
It  can be shown  that  the exact value of QFI density for this operator, for an arbitrary system size, is given by
\begin{align}
    &f_Q(\ket{\Phi_1})=f_Q(\ket{\Phi_1})=4\frac{3^M-1}{3^M+2+(-1)^M}, \label{eq:fQ_PBC} \\
    &f_Q(\ket{\Gamma_{\alpha,\beta}})=4\frac{(1-\frac{1}{M})3^M+1}{3^M+(-1)^{M+\alpha+\beta}}, \label{eq:fQ_OBC}
\end{align}
with $M=N/2$.
From Eqs.~\eqref{eq:fQ_PBC}-\eqref{eq:fQ_OBC} it is clear that regardless of the boundary conditions or the energy, the QFI density for these states asymptotically approaches 4 from  below. Numerical computations with other operators such as the total magnetisation led to the same intensive behaviour in system size, which is to be contrasted with the extensive scaling of the QFI density for the tower of $N{+}1$ eigenstates discussed in the main text. 

\end{document}